\documentstyle[epsfig,times]{l-aa}

\begin{document}
   \thesaurus{13.25.2 -- 11.19.1 -- 10.09.1}
   \title{X-ray detections of weak Seyfert~2's with BeppoSAX}

   \author{M. Salvati
          \inst{1}
   \and L. Bassani
          \inst{2}
   \and R. Della Ceca
          \inst{3}
   \and R. Maiolino
          \inst{1,} \inst{4}
   \and G. Matt
          \inst{5}
   \and G. Zamorani
          \inst{6}
          }

   \offprints{M. Salvati}

   \institute{Osservatorio Astrofisico
             di Arcetri, L.~E.~Fermi 5, I--50125 Firenze, Italy
   \and Istituto Tecnologie e Studio Radiazioni Extraterrestri, CNR, Via
              Gobetti 101, I--40129 Bologna, Italy
   \and Osservatorio Astronomico di Brera, Via Brera 28, I--20121 
             Milano, Italy
   \and MPI f\"ur Extraterrestrische Physik, Giessenbachstrasse 1,
             D--85748 Garching bei M\"unchen, Germany
   \and Dipartimento di Fisica, Universit\`a di Roma III, Via della Vasca
             Navale 84, I--00146 Roma, Italy
   \and Osservatorio Astronomico di Bologna, Via Zamboni 33, I--40126
             Bologna, Italy}

   \date{Received / Accepted }

   \maketitle

   \begin{abstract}

We report the detection in the X~rays of two weak Seyfert~2's 
(NGC3393 and NGC4941) with the
Italian--Dutch satellite BeppoSAX. These are among the first sources observed
in a sample of 12 Seyfert~2's which are being studied within the
BeppoSAX core program, in an effort to probe the putative torus at high
X-ray energies, calibrate an isotropic luminosity indicator for absorbed
nuclei, and determine the distribution of torus thicknesses, $N_H$.

Both a Compton thick spectrum, with a reflected power law and a large
equivalent width iron line, and a Compton thin spectrum, with the intrinsic
power law transmitted through a large column density absorber, provide
acceptable fits to both sources, with some preference for the latter model
in the case of NGC4941. The high initial detection rate in our program points 
to a large final X-ray sample.

   \end{abstract}

%

\section{Introduction}

A variety of observations have provided evidence that Seyfert~2 galaxies host 
a nuclear active source which is similar to that observed in Seyfert~1's,
but is obscured along our line of sight by a pc-scale
gaseous torus (see Antonucci 1993 for a review). The most elementary 
version of such a ``unified model'' assumes Seyfert~2's and Seyfert~1's 
to be identical physical objects, while the orientation of the line of 
sight with respect to the obscuring torus accounts for all the observed 
differences between the two classes. X-ray data have played a major role
among the observations supporting this model.
EXOSAT first, and Ginga later clearly proved that the
X-ray emission of several Sy2's is characterized by a power law
spectrum  similar to that observed in Sy1's (photon index $\sim$ 1.7),
with a cutoff at low
energies due to photoelectric absorption by gas of column density
between 10$^{22}$ and 10$^{24}$ cm$^{-2}$ (e.g. Awaki et al. 1991; Koyama 1992;
Nandra \& Pounds 1994). Such a dense absorbing column has been identified
with the torus.

Hard X-ray spectra were recognized as a powerful potential tool to test and
probe various issues on active nuclei. The direct component of the Sy2
X-ray emission (i.e. the emission at energies above the absorption cutoff) 
provides direct information on the intrinsic luminosity of the active nucleus; 
more specifically, it can be regarded as the only isotropic indicator of the
luminosity of Sy2 nuclei. The knowledge of the intrinsic luminosity should
allow to test whether Sy2 and Sy1 nuclei indeed have the same average power
(``strong'' version of the unified model), or there is some systematic
luminosity difference (Barcons et al. 1995; Falcke et al. 1995).
The distribution of absorbing column densities $N_H$ should provide
information on the geometry and physics of the obscuring material.
Eventually, by assessing the distribution of luminosities, spectral 
indices, and column densities
among Sy2's, the origin of the X-ray background should be tightly
constrained without assuming any specific model for the spectra of Sy2's 
(Setti \& Woltjer 1989; Comastri et al. 1995; Zdziarski et al. 1995; 
Barcons et al. 1995).

However, such goals have not been fully addressed yet. The available hard
X-ray spectra of Sy2's are affected by selection effects towards X-ray bright
sources, or sources known to be bright in either UV or IR.
Ginga surveys on Sy2's, such as Awaki et al. (1991), Koyama (1992) and
Mulchaey et al. (1992), contain a large fraction of Markarian galaxies
and/or sources bright in the HEAO1 survey. Therefore, such studies are not
representative of the full Sy2 population, but they rather sample the
high luminosity tail and/or the Sy2's absorbed by low column densities.
More recently, ASCA observations have improved the Sy2 census by
providing hard X-ray spectra of a larger number of sources.
As a result, Sy2's with lower luminosities and larger absorbing column 
densities have been detected (e.g. Ueno et al. 1996;
Awaki et al. 1996; Makishima et al. 1994). Yet, bias effects are still
significant and the statistics very limited. Also the paucity of highly
absorbed ($N_H > 10^{25} \rm cm^{-2}$), Compton thick Sy2's (only 
four candidates were known at the beginning of our program, see e.g. 
Matt 1997 and references therein) 
probably does not reflect the real abundance of such objects, but is 
rather a consequence of a selection effect,
as the X-ray spectrum of these sources is seen only in reflection and,
therefore, it is orders of magnitude fainter than in other Seyferts.

\section{The weak Seyfert~2 core program}

We have undertaken a program of observations with BeppoSAX, the 
Italian--Dutch X-ray satellite, 
aimed at properly assessing the distribution of luminosities
and absorbing column densities in a sample of Sy2's as little biased
as possible with respect to orientation effects. Objects suitable for such a 
survey were drawn from the Seyfert sample of Maiolino \& Rieke (1995). 
This sample is extracted from the
Revised Shapley Ames catalogue of galaxies, limited in total
blue magnitude (B$_T <$ 13.2), and Seyfert nuclei are identified
according to the properties of their (narrow) emission line spectrum.
Although, as discussed in Maiolino \& Rieke, the narrow line spectrum is
not a completely isotropic indicator of the nuclear activity, this Seyfert
sample is much less biased than others, both in terms of luminosity
of the nuclear source and of properties of the host galaxy.

There are 54 objects classified as Seyfert~2's in the Maio\-li\-no \& Rieke
sample; intermediate Seyferts could arise from a peculiar geometry (see below),
and we have not considered them. Out of the 54, 22 have already been observed
by ASCA, and for almost all of them hard X-ray information has been published 
(see Polletta et al. 1996 and references therein; see also the HEASARC
online database). We have prioritized the remaining targets in terms of their
[OIII] flux, under the assumption that the [OIII] emission is a sufficient 
approximation to an isotropic measure of the nuclear luminosity.
If the [OIII] emission were not completely isotropic, our initial
sample would be biased in favour of objects with lower optical absorption 
and, perhaps, lower values of $N_H$. The initial sample includes 12 
objects, to be observed during the first year of BeppoSAX; we aim at
the completion of the sample within the satellite lifetime; we are also
collecting optical data both from archives and from new observations.

According to the ``two--tori'' picture of Maiolino
\& Rieke (1995), the Seyfert nuclei are surrounded by a pc-scale torus,
oriented at random with respect to the host galaxy, and responsible for
the obscuration of the X~rays and of the broad lines, and by a larger and
less opaque screen ($N_H \sim 10^{22}$ cm$^{-2}$), 
coplanar with the galaxy, and responsible for the
intermediate Seyfert types. Within this scenario our initial sample
might be somewhat biased in favour of face-on Seyfert~2's, but would be free
from biases with respect to the inner, thicker torus. In any case, our data
would allow us to detect a possible bias by looking for an inverse 
correlation between $N_H$ and [OIII] at constant nuclear luminosity, i.e., at 
constant intrinsic $L_X$. Such a correlation would be important {\it per se},
since it would give information on the height of the torus;
and could be corrected for in order to restore the
usefulness of our results for the goals of the program.

\section{Data collection, analysis and results of spectral fitting}

The X-ray astronomy satellite BeppoSAX is a joint project of the Italian
Space Agency (ASI) and the Netherlands Agency for Aerospace Programs (NIVR)
developed by a consortium of institutes in Italy and The Netherlands,
including the Space Science Department of ESA (SSD). The scientific
payload comprises four Narrow Field Instruments [NFI: Low Energy
Concentrator Spectrometer (LECS), Medium Energy Concentrator Spectrometer
(MECS), High Pressure Gas Scintillation Proportional Counter (HPGSPC),
and Phoswich Detector System (PDS)], all pointing in the same direction,
and two Wide Field Cameras (WFC), pointing in diametrically opposed 
directions perpendicular to the NFI common axis.
A detailed description of the entire BeppoSAX mission can be found in 
Butler \& Scarsi (1990) and Boella et al. (1997a). The MECS 
(see below) consists of three equal units, 
with a field of view of 28~arcmin radius, working range 1.3--10~keV,
energy resolution $\sim$8\% and angular resolution $\sim$0.7~arcmin (FWHM) 
at 6~keV. The effective area at 6~keV is 155 cm$^2$ (Boella et al. 1997b).

Our targets were observed with the NFI for $\sim$15~ksec starting on January 
8.92 UT, 1997 (NGC3393), and for $\sim$29~ksec starting on January 22.18 UT, 
1997 (NGC4941). In both cases, only the MECS provided useful data, since the
LECS exposure was much shorter due to the switching off of the instrument
over the illuminated Earth, and the HPGSPC and PDS were not sensitive
enough for such low fluxes.
For each source, the three MECS were aligned, equalized, and summed; the
extraction radius was chosen equal to the suggested value of 4~arcmin;
the background was taken from much longer
exposures of the empty sky, and again extracted in a 4~arcmin radius 
centered approximately on the same detector coordinates as the source.
The net source counts are 104$\pm$16 (6.7 $\sigma$) and 307$\pm$24 
(13 $\sigma$) for NGC3393 and NGC4941, respectively.
After grouping the total counts in energy bins with a minimum of 20 counts 
each, and discarding any event outside the 2--10~keV interval because
of poor calibrations, we carried 
out the spectral analysis with XSPEC, complemented with the release 97.1
of the BeppoSAX response matrices. 

We have tried three spectral models, all including a gaussian line plus, 
respectively: a power law, a fraction of which is transmitted through cold 
matter (transmission model), a power law with zero absorption (scattering 
model), and a reflected spectrum due to the reprocessing of a power law by 
cold matter (reflection model). The torus surrounding the nucleus is assumed 
to be Compton thin in the transmission scenario, Compton thick in the 
remaining cases; in these latter cases a mirror is required, which can
be either grey (scattering) or energy dependent (reflection). 

In the case of NGC3393 the small number of bins does not allow an 
elaborate modeling. In order to cope with the
low statistics we have frozen to standard values as many parameters as
possible; in particular, we have kept the intrinsic spectral index
at 1.7, which is the ``universal'' value outside a few Schwarzschild
radii. The scattering model gives an acceptable $\chi^2$ (6.1 with 8 d.o.f.),
but requires a hot, ionized plasma for the mirror; if we assume that all
iron is H--like or He--like, we find that the equivalent width of the
line, EW, is larger than 4.5~keV at the 90\% level; such a value is
incompatible with cosmic abundances for all but the lowest column densities
(Matt et al. 1996). The best fit parameters of the remaining models ($\chi^2$ 
of 2.1 and 3.8 with 6 and 7 d.o.f., respectively) are shown in Table~1.
\begin{table}
\caption{Spectral fit parameters for NGC3393, for transmission
and reflection models, respectively. Frozen parameters are: photon 
indices (1.7), line energy (6.4~keV), line width (0.1~keV).
Normalizations are in photons/cm$^2$/s/keV at 1~keV for power laws, and in 
photons/cm$^2$/s for lines; line EW are in keV, and are calculated with
respect to the observed continuum; column densities
are in units of $10^{22}$~atoms/cm$^2$; model fluxes are in units of
$10^{-13}$~erg/cm$^2$/s in the 2--10~keV range; luminosities in 10$^{41}$ erg/s
in the 2--10 keV range, after correction for absorption and assuming $H_0$=50
km/s/Mpc; the confidence intervals
are 90\% with one interesting parameter.}

\vglue0.5truecm
{\begin{tabular}{l l l}
Model       & Parameter Name      & Value and Range  \\
&&\\
\hline
&&\\
Transmission      & Normalization \#1   & 5.05~(0.43--213)  $\times 10^{-4}$ \\  
             & Column Density      & 77.5~($>$~17.0) \\
             & Normalization \#2   & 4.65~(1.61--7.19) $\times 10^{-5}$ \\
             & Line Normalization  & 9.07~(0.89--17.1) $\times 10^{-6}$ \\ 
             & Line EW             & 1.18~(0.08-3.35) \\
             & Model Flux          & $5.57\pm0.83$      \\
             & Luminosity          & $16.8\pm2.5$      \\
&&\\
\hline
&&\\
Reflection      & Normalization       & 1.75~(0.04--3.73) $\times 10^{-3}$ \\
             & Escape fraction     & 0.02~(0.00--1.00) \\
             & Line Normalization  & 1.18~(0.32--1.88) $\times 10^{-5}$ \\
             & Line EW             & 2.08~(0.42--5.03)              \\
             & Model Flux          & $4.99\pm0.75$    \\
             & Luminosity          & $4.05\pm0.60$      \\
&&\\
\end{tabular}}
\end{table}
Both models are acceptable, from a statistical and an astrophysical point 
of view, and indeed they overinterpret the data.
If the reflection scenario is correct, the intrinsic luminosity of the
source should be at least an order of magnitude greater than the observed
one, 4.1$\times10^{41}$ erg s$^{-1}$, more in line with the [OIII]
luminosity.

In the case of NGC4941, a scattering model gives
a $\chi^2$ of 25.5 with 21 d.o.f., and a {\it negative} photon 
index, which is obviously an artifact of a strongly absorbed component. 
If the photon index is frozen at 1.7, the model can be rejected with
a $\chi^2$ of 47.8 for 22 d.o.f. 
The remaining models are both formally acceptable ($\chi^2$ of 16.6 and
23.5 with 20 and 21 d.o.f. for the transmission and the reflection
scenario, respectively), but the latter does not reproduce adequately
the lowest energy points. We have thus tried to ignore this low energy
excess (which could be due to a different source, say a starburst), and
to fit the remaining points on their own. The transmission scenario
is still marginally better than the reflection one, in that it
reduces the $\chi^2$ by more than 2.71 (the 90\% confidence limit)
at the expense of one additional parameter; the difference lies in
the roll over observed below $\sim$5~keV, which is difficult to fit
with a reflected power law of intrinsic slope 1.7. Table~2 gives the
best fit parameters of the two models; in particular, the iron line EW
is within the expected ranges in both cases (Ghisellini et al. 1994;
Matt et al. 1996). In the transmission 
scenario the luminosity (after correction for absorption) is 
2.1$\times$10$^{41}$ erg s$^{-1}$, which places NGC4941 at the lower end
of the Seyfert luminosity function. 

\begin{table}
\caption{The same as Table~1 for NGC4941.}
\vglue0.5truecm
{\begin{tabular}{l l l}
Model       & Parameter Name      & Value and Range  \\
&&\\
\hline
&&\\
Transmission      & Normalization \#1   & 8.37~(4.18--20.5) $\times10^{-4}$ \\
             & Column Density      & 69.1~($>$~40.6) \\
             & Normalization \#2   & 4.85~(2.78--6.51) $\times10^{-5}$  \\
             & Line Normalization  & 1.27~(0.64--1.83) $\times10^{-5}$ \\
             & Line EW             & 0.98~(0.43--1.53)      \\
             & Model Flux          & $8.54\pm0.67$        \\
             & Luminosity          & $2.06\pm0.16$        \\
&&\\
\hline
&&\\
Reflection      & Normalization       & 4.27~(3.34--5.20) $\times 10^{-3}$ \\
($>$~3~keV)  & Escape fraction     & 0.00~(frozen) \\
             & Line Normalization  & 1.49~(0.87--2.00) $\times 10^{-5}$ \\
             & Line EW             & 1.41~(0.77--2.07)              \\
             & Model Flux          & $7.54\pm0.59$    \\
             & Luminosity          & $0.453\pm0.036$      \\

\end{tabular}}
\end{table}

The observed spectra of the two galaxies with the best fitting models 
superposed on them are given in Figs.~1 and 2 for NGC3393 (reflection 
scenario) and NGC4941 (transmission scenario), respectively.

\begin{figure}
\epsfig{file=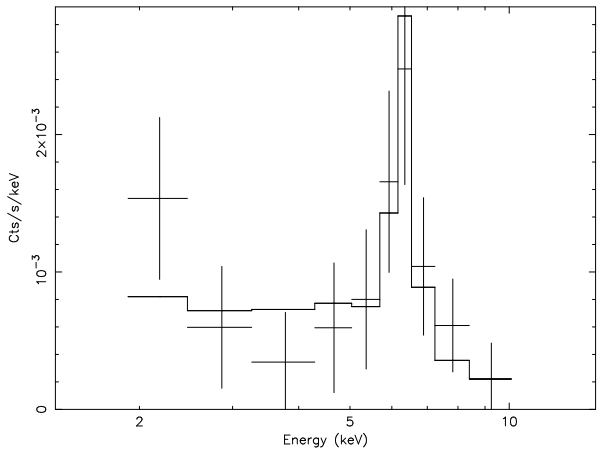, height=6.7cm, angle=-90, silent=}
\caption{The observed BeppoSAX spectrum of NGC3393 with the reflection
best fit model. The fit parameters are given in Table~1.}
\end{figure}

\begin{figure}
\epsfig{file=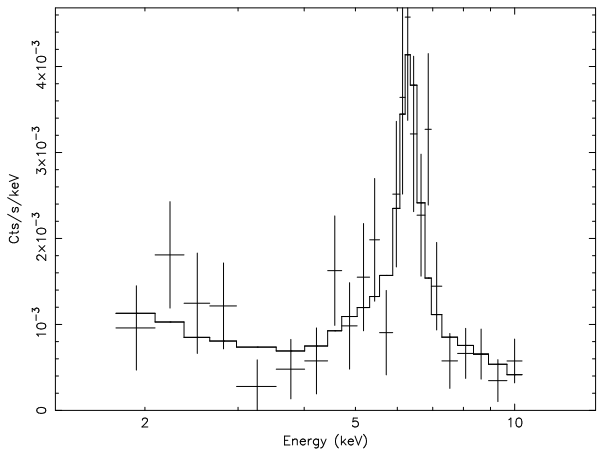, height=6.7cm, angle=-90, silent=}
\caption{The observed BeppoSAX spectrum of NGC4941 with the transmission
best fit model. The fit parameters are given in Table~2.}
\end{figure}

\section{Discussion and conclusions}

While a firmer basis for discussion will be provided only by the completion
of the program, it is tempting to put forward a few comments. First of all,
for the first targets we have a detection rate of 100\% 
(at the time of writing five further observations were performed, and they too
resulted in detections): this is at least a suggestion that at the level of 
a few $\times 10^{-13}$ erg/cm$^2$/s nearby Seyfert~2's are frequently visible
at medium--hard X-ray energies. 

More troublesome appear the prospects of measuring the column density
$N_H$ with a reasonable precision; while we are certain that NGC3393 and
NGC4941 both have $N_H$ larger than several times $10^{23}$~cm$^{-2}$,
with the X-ray data alone and with the available statistics we cannot
discriminate between a Compton thin or a Compton thick torus. This
uncertainty entails a large uncertainty on the intrinsic X-ray luminosity.
Previous studies have pointed to the far infrared luminosity (FIR) or
to the [OIII] luminosity as isotropic indicators of the ``true'' 
luminosity of the nucleus (e.g. Mulchaey et al. 1994; Awaki 1997).
In our case, assuming that Sy1's and Sy2's are intrinsically identical,
assuming that $L_X/L_{FIR}=0.2$ and $L_X/L_{[OIII]}=100$ for Sy1's
(Mulchaey et al. 1994), and \hbox{--finally--} assuming that both galaxies
have Compton thin tori, we find that NGC4941 is underluminous in X~rays
by a factor of $\sim10$, and NGC3393 by $\geq40$. On the other hand,
assuming that both galaxies have Compton thick tori, and we see only 1\%
of the intrinsic flux as reflected flux, we find reasonable agreement 
with the Sy1 normalization. 

With many more data points of similar quality we will be able to check
the self--consistency of the unified model. For an independent test of it,
however, we would have to measure $N_H$ and the intrinsic $L_X$ without
a priori assumptions; this would require longer exposures, in order
to acquire better statistics in the 2--10~keV range with the MECS, and to 
obtain detections or significant upper limits at energies $\geq$~20~keV with
the PDS, where in the reflection scenario a Compton hump would be 
expected.

\begin{acknowledgements}
We are grateful to all the people who at all levels of responsibility
have worked over the years at the BeppoSAX project. We thank in particular
Fabrizio Fiore, for his help with the detector responses.
This research has made use of SAXDAS linearized and cleaned event 
files (Rev0) produced at the BeppoSAX Science Data Center. Two of us 
(M.S. and R.M.) acknowledge the support of the ASI grant ASI--95--RS--120.
\end{acknowledgements}

\end{document}